\pdfoutput=1

\documentclass[acmtog]{acmart}
\acmSubmissionID{440}
\usepackage{bm}
\usepackage[ruled]{algorithm2e} 

\SetAlFnt{\small}
\SetAlCapFnt{\small}
\SetAlCapNameFnt{\small}
\SetAlCapHSkip{0pt}

\AtBeginDocument{%
  }

\setcopyright{acmlicensed}
\acmYear{2023}

\acmJournal{TOG}
\acmMonth{12}

\begin{document}

\title{DROP: Dynamics Responses from Human Motion Prior and Projective Dynamics}

\author{Yifeng Jiang}
\affiliation{%
  \institution{Stanford University}
  \country{United States of America}
}
\email{yifengj@stanford.edu}

\author{Jungdam Won}
\affiliation{%
  \institution{Seoul National University}
  \country{South Korea}
}
\email{jungdam@imo.snu.ac.kr}

\author{Yuting Ye}
\affiliation{%
  \institution{Meta Reality Labs Research}
  \country{United States of America}
}
\email{yuting.ye@meta.com}

\author{C. Karen Liu}
\affiliation{%
  \institution{Stanford University}
  \country{United States of America}
}
\email{karenliu@cs.stanford.edu}

\begin{teaserfigure}
\centering
\includegraphics[width=\textwidth]{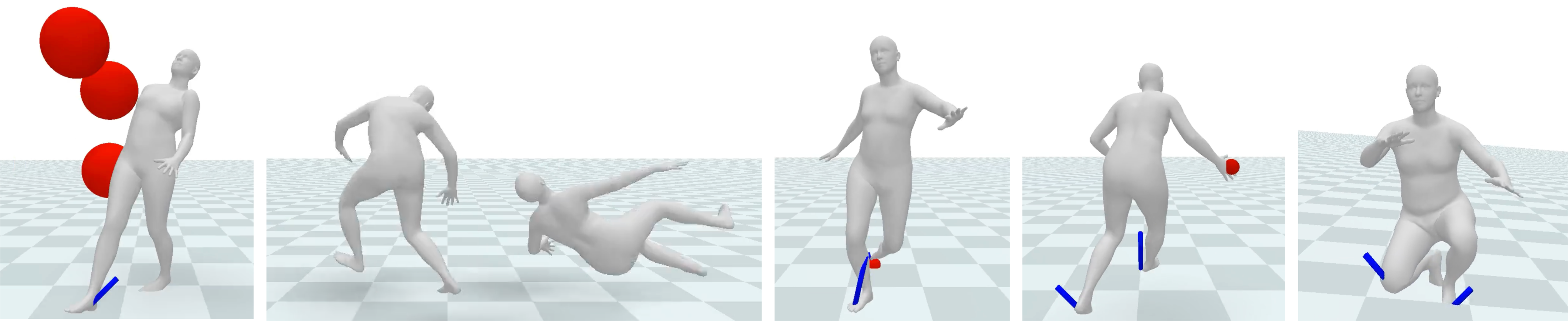}   
\caption{DROP synthesizes dynamic reaction and recovery motion in response to a variety of perturbations. From left to right, responses to projectiles, two-character collision, tripped by an obstacle, target following, and responses to a tilting platform.}
\label{fig:teaser}  
\end{teaserfigure}



\begin{abstract}
  Synthesizing realistic human movements, dynamically responsive to the environment, is a long-standing objective in character animation, with applications in computer vision, sports, and healthcare, for motion prediction and data augmentation. Recent kinematics-based generative motion models offer impressive scalability in modeling extensive motion data, albeit without an interface to reason about and interact with physics. While simulator-in-the-loop learning approaches enable highly physically realistic behaviors, the challenges in training often affect scalability and adoption. We introduce DROP, a novel framework for modeling Dynamics Responses of humans using generative mOtion prior and Projective dynamics. DROP can be viewed as a highly stable, minimalist physics-based human simulator that interfaces with a kinematics-based generative motion prior. Utilizing projective dynamics, DROP allows flexible and simple integration of the learned motion prior as one of the projective energies, seamlessly incorporating control provided by the motion prior with Newtonian dynamics. Serving as a model-agnostic plug-in, DROP enables us to fully leverage recent advances in generative motion models for physics-based motion synthesis. We conduct extensive evaluations of our model across different motion tasks and various physical perturbations, demonstrating the scalability and diversity of responses.
\end{abstract}

\ccsdesc[500]{Computing methodologies~Animation}
\ccsdesc[500]{Computing methodologies~Physical simulation}

\keywords{\textbf{Generative Models, Physics-based Motion Synthesis, Hybrid Methods}}


\maketitle

\newcommand{\norm}[1]{\left\lVert#1\right\rVert}
\newcommand{\argmin}{\operatornamewithlimits{argmin}}
\newcommand{\sign}{\operatornamewithlimits{sign}}


\section{Introduction}

Creating virtual humans that can move autonomously and realistically stands as a key pursuit in character animation. Crucial to this endeavor is endowing these virtual characters with the ability to adapt to and dynamically interact with the environment. Characters imbued with such capabilities can enhance player immersion in games and AR/VR environments, provide a predictive human model for computer vision tasks, offer a safe and cost-efficient means to synthesize or augment data for robotics applications, and support healthcare and sports applications.


Recent advancements in character animation indicate that generative models, such as Variational Autoencoders (VAEs) or Diffusion Models (\cite{rempe2021humor, ling2020character, tevet2023human}), can efficiently learn human motion priors from large-scale mocap datasets. However, these learned kinematic models (motion priors) do not incorporate physics thus lack responses to perturbations in the environment. Previous attempts to incorporate physics to motion priors require either additional training of control policies using reinforcement or imitation learning, or re-training of motion priors with some notion of physics. Both approaches inevitably increase design and computational complexity, potentially hindering adoption and scalability of the methods. In contrast, this paper investigate the possibility to decouple the incorporation of physics from the training of kinematic models, exploiting scalability and modularity inherent from pre-trained generative models.

We introduce DROP, a motion-model-agnostic plug-in human physics simulator that adds dynamics capability to a pre-trained kinematic generative model, without the need to retrain the motion prior model or train another policy to control it. DROP simulates quasi-physics effects via an optimization-based implicit Euler integrator with projective dynamics, and incorporates them with an existing kinematic model (we use HuMoR \cite{rempe2021humor}). Our insight is that generative kinematic models inherently possess a capacity to generalize beyond the training distribution, particularly in recovering from out-of-distribution states, due to their design and training methodologies. However, unlike the recovery capabilities of simulator-in-the-loop methods, such as reinforcement learning, their recovery trajectories are oblivious to physics, making them often implausible and unreliable. Therefore, we develop a minimalist, physically plausible human simulator to collaborate with any autoregressive motion prior, directing its recovery trajectory towards physical validity, with no additional training. 


Crucially, we employ projective dynamics \cite{Bouaziz2014Projective}, an optimization-based (variational) simulation framework, to simulate our character which is modeled as a rigid stick figure aligning with the state representation of many kinematic models. Because an autoregressive motion prior provides a state-dependent manifold that models the probability distribution of the next state, we can seamlessly integrate the motion prior as one of the energy fields in the projective dynamics framework. It collaborates with other physics-based energy terms, such as contact and rigidity, to balance the equation of motion in the framework of variational integration. Hence, our work can also be viewed as an extension of example-based simulation \cite{Martin2011Example, Jones2016Example}, where the example manifold now represents state transitions instead of poses.


Our method opens avenues for various downstream applications that utilize motion priors, with minimal modifications and no further training. Changes to the physical properties of the skeleton, new environmental constraints, path or target tracking, can each be formulated as an additional energy term, while preserving existing projective energies. In our experiments, we present different plug-and-play examples to showcase the flexibility of our framework. Furthermore, we demonstrate that our model can generate diverse responses to environmental forces across various motion categories. Quantitative measures and ablation studies show that our method effectively regularizes the recovery of motion from perturbed, out-of-distribution states.

\section{Related Work}


We review existing research in motion modeling, dynamic motion simulation, and hybrid methods that incorporate physics simulation with motion models. 


\subsection{Kinematic Motion Generation}

Creating fluid and continuous motions for virtual characters is a core challenge in animation. The seminal work of Motion Graph \cite{Kovar2002Motion, Arikan2002Interactive, Lee2002Interactive} arranges pre-recorded motion segments into a graph that can be traversed in runtime to produce seamless and extended motions. Although the results are high quality, the sparse graph connections limit responsiveness. Subsequent methods such as Motion Fields \cite{Lee2010Motion} and Motion Matching \cite{Clavet2016Motion} instead leverage k-nearest neighbors to determine the next pose for any given state, enabling instant adaptation to changes. These search-based algorithms unfortunately face difficulties when scaling to large, diverse, but unbalanced datasets such as AMASS \cite{AMASS:ICCV:2019}.

Recent approaches use deep networks to model motion transitions from large datasets. Neural networks can accelerate the motion matching process \cite{Holden2020Learned}, or automatically learn the best features for specific downstream tasks through regression models and supervised learning \cite{Holden2017PFNN, Starke2019Neural, starke2022DeepPhase}. Generative models, such as (Conditional) Variational Autoencoders (VAE, CVAE) \cite{ling2020character,rempe2021humor}, Generative Adversarial Networks (GAN) \cite{MEN2022GAN, Barsoum2018HPGAN}, or Diffusion Models \cite{tevet2023human}, focus on learning the distribution of motion transitions to better represent the inherent variability in human motions. These models can generate the next state based on the predicted current state through sampling. In our work, we utilize a pre-trained generative model, HuMoR, but we enforce physical realism in the generated transitions and use projective dynamics to minimize deviation from the learned distribution. Our approach can synthesize novel connections amongst latent states, avoid divergence to unstable states, and guide the recovery back to the learned distribution when encountering unseen perturbations.

\subsection{Dynamic Motion Simulation} 

Contrary to motion models that reproduce motions similar to those in the dataset, physics-based motion control naturally yields realistic dynamics responses to forces and changes in the environment. Techniques using optimization \cite{Ye:2010:OFC, daSilva2008Interactive, Muico2009Contact} and reinforcement learning \cite{Peng2018DeepMimic, Liu2017Learning} have proven effective in simulating a single example sequence under moderate perturbations. The source of these examples can be motion capture, hand animation, or outputs from motion synthesis \cite{Bergamin2019DReCon, park2019learning}. To expand the controller from a single example to a diverse motion dataset, methods akin to generative motion models are adopted in control policies. These include employing Motion Graphs with mixture-of-expert networks \cite{Won2020Scalable}, supervised policy learning with differentiable simulation \cite{supertrack}, using Conditional Variational Autoencoders (CVAEs) \cite{Won2022Physics, Yao2022ControlVAE}, or using Generative Adversarial Networks (GANs) \cite{Peng2021AMP, Peng2022ASE, iccgan2021}. However, the complexity of the training process and computation time also scale with the dataset diversity. Additionally, these control policies tend to produce stiff physical responses \cite{Xie-2023-PolicyEval}, and creating human-like compliance may require tuning simulation parameters and modifying the training process for each different downstream task \cite{Lee2022Deep}. For realistic recovery behavior, specific controllers can be learned from captured recovery motions \cite{Shiratori2009Simulating}. In contrast, our method serves as a decoupled add-on module to a generic pre-trained kinematic motion model, and does not require any task-specific learning of the motion model or a control policy.

\subsection{Hybrid Simulation and Kinematic Methods}

Closest to our method are techniques that enhance kinematic motions with physical and dynamics responses at the moment of interaction. Zordan and his colleagues \shortcite{Zordan2002Motion, zordan2005dynamic} blend simulated passive motions with kinematic motions at the onset and offset of interaction. It produces convincing responses but the transitions are not always realistic. Ye and Liu \shortcite{Ye:2008:ARC} apply passive simulation in a torque space that does not interfere with the input motion, but their method cannot handle balance. Example data of people being pushed can be adapted to runtime perturbations. Arikan et al. \shortcite{Arikan2005Pushing} use simulated motions to retrieve and modify the closest example. Similarly, Ye and Liu \shortcite{Ye2010Synthesis} learn a latent dynamic model from perturbation and recovery examples, and apply simulation to search in the latent space of recovery motions. Motion Field \cite{Lee2010Motion} also provides an add-on method for perturbation recovery. While effective, these methods are restricted by situations that can be captured safely and realistically. They also do not scale well to diverse dynamic situations. In comparison, our method does not require the motion dataset to contain examples of people being pushed or falling over, provided enough diversity in the data and we can effectively identify and connect relevant spaces within the data during interactions. Our method also does not switch in and out of simulation.

\section{Algorithm Overview}


We introduce a human motion simulator based on projective dynamics, an optimization-based (variational) implicit Euler integrator. Projective dynamics provides a flexible interface to incorporate the generative motion prior as a projective energy term. Optimization-based simulation provides stability under large time steps, which facilitates integrating existing motion priors operating at arbitrary frequencies. Using this framework, we first create a minimalist uncontrolled ragdoll model, adhering to basic human kinematic and dynamic invariants, such as contact, rigidity, and range of motion (RoM), with each invariant represented as an energy for the variational integrator. 

The learned motion prior then serves as an additional energy term, effectively supplying control forces to the character. The choice of projective dynamics over other optimization integrator variants is crucial, as it permits the incorporation of any class of autoregressive generative motion prior, so long as its output can be represented as a state-dependent manifold. For a more detailed discussion on this design choice, see Sec. \ref{sec:alternative}.


Further, we introduce a simple and effective correction technique to reduce the use of "magic force" from the motion prior energy term by regulating the linear and angular momentum of the center-of-mass (CoM).

\section{Minimal Character Simulation with Projective Dynamics} 

We begin by constructing an uncontrolled ragdoll simulator.
As shown in Fig. \ref{fig:method}, our simulation character consists of $N$ mass particles, each positioned at a human joint (excluding hands, toes, and head) with rest-pose locations provided by the SMPL human model with an average body shape \cite{SMPL:2015}. Massless sticks connect the joint particles, modeled as strong springs similar to recent work \cite{Chen2022Unified} that extends the IPC \cite{Li2020IPC} framework to articulated rigid-bodies. Consequently, the state $\bm{x}$ of our simulated ragdoll can be fully determined by joint particle locations, with $\bm{x} \in \mathbb{R}^{3N}$ and $N = 22$ according to SMPL.

Using an optimization-based (or variational) integrator \cite{liu2013fast, gast2015optimization, Martin2011Example}, we solve for the next state $\bm{x}_{t+1}, \bm{v}_{t+1}$ from $\bm{x}_{t}, \bm{v}_{t}$, and external forces $\bm{f}_{t}$ by:

\begin{eqnarray}
    \bm{x}_{t+1} &=& \argmin_{\bm{x}} \frac{1}{2h^2} (\bm{x}-\bm{y})^T \bm{M} (\bm{x}-\bm{y}) + \sum_i E_i(\bm{x}; \bm{x}_t, \bm{v}_t), \label{eq:dyn1} \\
    \bm{v}_{t+1} &=& (\bm{x}_{t+1} - \bm{x}_t) / h,
\end{eqnarray}
where  $\bm{y} = \bm{x}_{t} + h \bm{v}_t + h^2 \bm{M}^{-1} \bm{f}_{t} + h^2 \bm{g}$, $\bm{g}$ represents the gravity vector, and $h$ is the simulation step length (30Hz in this work to match the learned kinematics motion prior). The first term in Eq. \ref{eq:dyn1}, $E_{mom} = \frac{1}{2h^2} (\bm{x}-\bm{y})^T \bm{M} (\bm{x}-\bm{y})$, arises from integrating backward (implicit) Euler discretization over $\bm{x}$. This "momentum" energy term holds second order (Newtonian) dynamics, allowing environment to change the system states through forces $\bm{f}_t$. The $\{E_i\}$ comprise energy terms for contact, RoM, and rigidity constraints, representing invariants we want to adhere to regardless of whether our character is controlled.



The projective dynamics method alternates between slack variable projection and solving steps at each optimization iteration. The projection step independently constructs a slack variable for each $E_i$ in Eq. \ref{eq:dyn1} (termed "local" step), while the solving step considers all $\{E_i\}$ collectively (termed "global" step). In practice, the local projection step can employ any routine, even non-differentiable ones like sampling, which provides flexibility for our framework. 


In the local step, we independently construct a manifold $\mathcal{C}_i$ for each $E_i$ and define $E_i$ as \cite{Bouaziz2014Projective}:

\begin{equation}
    \label{eq:manifold}
    E_i(\bm{x}) = \frac{w_i}{2} \min_{\bm{p}_i \in  \mathcal{C}_i}\norm{\bm{A}_i x - \bm{p}_i}^2,
\end{equation}

where $\mathcal{C}_i$ is the manifold when $\bm{A}_i x \in \mathcal{C}_i$ we have $E_i(\bm{x}) = 0$. Eq. (\ref{eq:manifold}) requires solving the slack variable $\bm{p}_i$ by projecting $\bm{A}_i x$ onto $\mathcal{C}_i$. For example, consider the case where $\bm{A}_i = I$, and $\mathcal{C}_i$ contains only one constant point $\bm{\bar{x}}$. In this case, we can easily obtain the slack variable $\bm{p}_i$ to always be $\bm{p}_i = \bm{\bar{x}}$. Consequently, $E_i(\bm{x}) = \norm{\bm{x} - \bm{\bar{x}}}^2$ essentially creates a spring force ($\nabla E$) that continually attracts the optimization variable $\bm{x}$ to the single-point manifold $\bar{\bm{x}}$.

In the global step, treating all solved $\bm{p}_i$'s as constants, $\sum_i E_i$ becomes a quadratic function. Finding the minimizer $\bm{x}^*$ $=$ $\argmin$ $(E_{mom} + \sum_i E_i)$ requires solving the linear system of equations:

\begin{equation}
\label{eq:linear-solve}
    (\bm{M}/h^2 + \sum_i w_i \bm{A}_i^T \bm{A}_i) x = \bm{M}\bm{y}/h^2 + \sum_i w_i \bm{A}_i^T \bm{p}_i
\end{equation}
obtained from $\nabla E = 0$.

\subsection{The Energy Terms}


\paragraph{Rigidity} Following previous works, we approximate rigid bones \cite{Chen2022Unified} with strong springs \cite{liu2013fast}:
\begin{equation}
E_1 = \frac{w_1}{2} \min_{\bm{d} \in  \mathcal{C}_1} ( \bm{x}^T \bm{L} \bm{x} - \bm{x}^T \bm{J} \bm{d} ),
\end{equation}
where $\mathcal{C}_1$ is the manifold of rest-length spring directions, and $\bm{d}$ is the vectors of springs projected back to their rest length. For a detailed derivation of $\bm{L}$ and $\bm{J}$, we direct readers to \cite{liu2013fast}.

\paragraph{Contact} To mitigate penetration, we utilize a simplistic contact model:
\begin{equation}
E_2 = \frac{w_2}{2} \min_{\bm{x_c} \in  \mathcal{C}_2} (\bm{x}^T \bm{S} \bm{x} - \bm{x}^T \bm{S} \bm{x}_c),
\end{equation}
where $\bm{S} \in \mathbb{R}^{3N \times 3N}$ is a diagonal selector matrix. Joint particles $x^{j}, j = 1, \cdots, N$ that will be in collision with the environment trigger corresponding entries in $\bm{S}$ to 1. $\bm{x}_c$ is the projection of the current configuration $\bm{x}_t$, along contact normal directions, back to the nearest penetration-free configuration. Consequently, $E_2$ effectively prompts all particles that are on the verge of penetration to exit the collision boundary in the normal direction, yet remain static in the tangential direction. There is a static friction assumption commonly seen in human motion synthesis. While integrating a more advanced IPC-style penetration-free contact model \cite{lan2022pdipc} should improve our framework, we leave that for future work.

\paragraph{Range of Motion (RoM)} We introduce another energy term to regulate the full body pose to remain within the joint ranges of motion:
\begin{equation}
\label{eq:rom-energy}
E_3 = \frac{w_3}{2} \min_{\bm{x_r} \in  \mathcal{C}_3} \norm{\bm{x} - \bm{x}_r}^2,
\end{equation}
Here, $\bm{x}_r$ is the projection of the current pose back to the nearest pose within joint limits, with the routine involving a neural-net model, VPoser \cite{SMPL-X:2019}. We provide further details in Sec. \ref{sec:impl}.

\section{Casting Motion Prior as Projective Energy}
\label{sec:motion-prior}

The kinematic motion prior serves as the "control" of the human character. The straightforward alternative for incorporating a motion prior is to train a model from massive motion data in the form of $E_{kin}(\bm{\tilde{x}}_{t+1}, \bm{x}_{t}, \bm{x}_{t-1}) \in \mathbb{R}$ (or equivalently $E_{kin}(\bm{\tilde{x}}_{t+1}, \bm{x}_{t}, \bm{v}_{t})$), where a realistic transition result in a low energy value and high energy vice versa, attracting the system state $\bm{x}_{t+1}$ back to the natural motion manifold. Training such differentiable energy functions as Energy-Based Models (EBM) \cite{lecun2006tutorial} is possible, but we opt for projective dynamics which allows us to directly take an existing motion prior not necessarily in the form of an EBM, without the need for retraining.


A general motion prior model typically takes the following form:
\begin{equation}
    \bm{\tilde{x}}_{t+1} = \mathcal{M} (\bm{x}_{t}, \bm{x}_{t-1}, ..., \bm{z}).
\end{equation}
This model aims to generate the next motion frame given the current state, an optional short motion history, and a noise vector $\bm{z}$ sampled from a predefined prior distribution (e.g., Gaussian). $\mathcal{M}$ is often a deep neural network. The model is considered "generative" since different sequences of $\bm{z}$ vectors would, over time steps, produce various motions, capturing the variability in natural human motion. We only consider \textit{autoregressive} motion models, which predict one frame at a time, as we build online dynamic simulation and response, where changes in the environment (e.g., force perturbation) are unknown beforehand.

A generative model in such a form provides a straightforward way to approximate its transition manifold: if we sample a set of ${\hat{\bm{z}}}$, we can estimate the transition manifold of $\mathcal{X}_{t+1}$ conditioned on known history frames with the set of ${\hat{\bm{x}}_{t+1}}$ samples:

\begin{equation}
    \mathcal{\hat{X}}_{t+1} := \{\hat{\bm{x}}_{t+1} | \hat{\bm{x}}_{t+1} = \mathcal{M} (\bm{x}_{t}, \bm{x}_{t-1}, ..., \bm{\hat{z}}) ~{\rm and}~ \hat{\bm{z}} \sim \mathcal{N}_0 \},
\end{equation}
where $\mathcal{N}_0$ is the prior distribution (we denote $\bm{\mu}_0$ as its mode). This then naturally fits into the projective dynamics framework with $E_{kin}$ defined as:
\begin{equation}
     E_{kin} = \frac{w_{kin}}{2} \min_{\bm{x}_{kin} \in \mathcal{X}_{t+1} } || \bm{x} - \bm{x}_{kin} ||^2,
\end{equation}
where we obtain $\bm{x}_{kin}$ in the local step by simply searching in $\hat{\mathcal{X}}_{t+1}$ the nearest point to the current solver state $\bm{x}$.

\begin{figure}[!t]
    \centering
    \includegraphics[width=\linewidth]{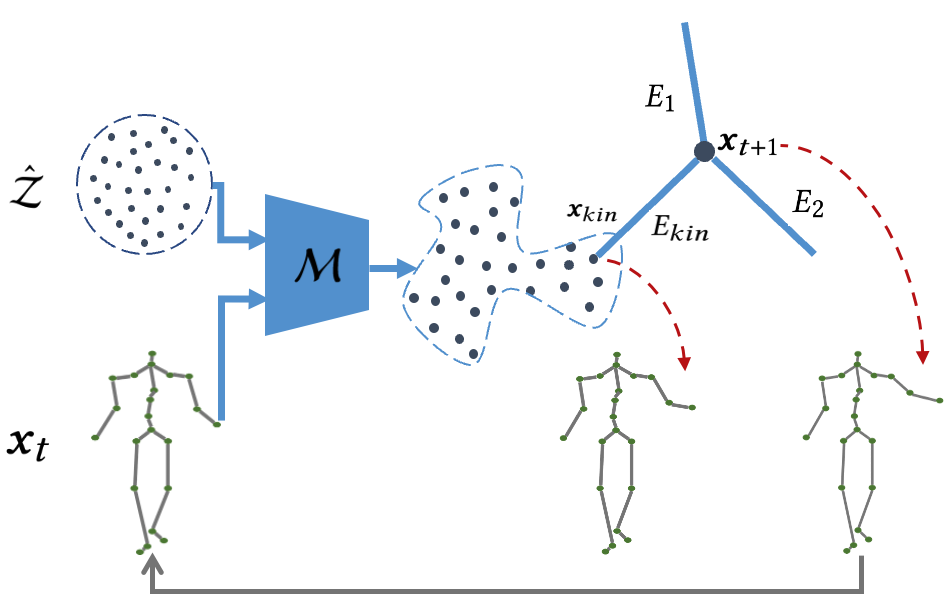}
    \caption{DROP: casting motion prior as one of the projective energies. Physics-based energies $\{E_i\}$ drive the simulated character with projective dynamics. Given a motion prior (e.g. conditional VAE decoder), an additional data-driven energy $E_{kin}$ is incorporated with the slack variable $x_{kin}$ being the projection of the solver state $\bm{x}$ onto the motion prior manifold.}
    \label{fig:method}
\end{figure}

In conjunction with the other physics-based energy terms, we present the main simulation loop (Algorithm \ref{alg:sim_loop}, Fig. \ref{fig:method}). At each simulation time step, we first batch-sample the motion prior to estimate $\hat{\mathcal{X}}_{t+1}$. Subsequently, we execute a few (3, in our implementation) projective dynamics iterations, where we alternate between solving for the projection slack variables $\bm{p}_i$ for each physics-based $E_i$, searching for $\bm{x}_{kin}$ for $E_{kin}$, and then finding the minimizer of (now quadratic) $E_{mom} + E_{kin} + \sum_i E_i$ with the solved $\bm{p}_i$'s and $\bm{x}_{kin}$ treated as known constants.


The success of our method depends on two hypotheses on the motion prior. First, $E_{kin}$ from the motion prior should roughly agree with other energies $\{E_i\}$ when the transitions are in distribution of the training data. That is, they should all be small when simulating an in-distribution, unperturbed motion. This assumption is reasonable if we design the $\{E_i\}$ according to real-world physics and the motion prior is learned from real-world data. 

Second, the motion prior has certain level of inherent ability to return to training distribution from a state that is out-of-distribution, albeit via nonphysical paths due to no data coverage. 
This assumption can be achieved because most autoregressive motion priors are trained by methods that enhance robustness, preventing error accumulation from causing significant drifts over long-sequence generation. For instance, MotionVAE \cite{ling2020character} demonstrated that scheduled sampling \cite{bengio2015scheduled} during training allows the model to recover to the training distribution over time. Likewise, robustness is inherent in Diffusion Models due to the large noise added during training. 

Under these two assumptions, the main task of physics-based energies $\{E_i\}$ would be regulating the recovery path to be more physically valid. This is achieved by directing the solved state $\bm{x}$ to compromise between $E_{kin}$ and $\{E_i\}$, as well as the momentum correction method introduced in the next section.


\begin{algorithm}
\caption{DROP Simulation Step}\label{alg:sim_loop}
{
    Input: ${\rm Motion\;Prior}$ $\mathcal{M}$, ${\rm Current\; State}$ $\bm{x}_{t},\bm{x}_{t-1}$ \\
    $\hat{\mathcal{Z}} \sim \mathcal{N}_0$ \\
    $\mathcal{\hat{X}}_{t+1} = \mathcal{M} (\bm{x}_{t}, \bm{x}_{t-1}, \hat{\mathcal{Z}})$, 
    $\bm{x}_{t+1} = \mathcal{M} (\bm{x}_{t}, \bm{x}_{t-1}, \mu_0)$ \\
    \For{$l=0,1,\cdots,{\rm numProjectiveDynamicsIter}$}
    {
       $  \bm{x}_{kin} = \argmin_{\bm{x} \in \hat{\mathcal{X}}_{t+1}} || \bm{x} - \bm{x}_{t+1} ||^2 $ \\
       $E_{kin} = \frac{w_{kin}}{2} || \bm{x}_{t+1} - \bm{x}_{kin} ||^2$  // \textrm{Note,~} $E_{kin}=0 \textrm{~when~} l=0$ \\
       Routines to calculate other projection variables $\bm{d}$, $\bm{x}_c$, $\bm{x}_r$ \\
       Solve $\bm{x}_{t+1} = \argmin E = \argmin [E_{mom} + E_{kin} + \sum_i E_i]$ \\
       
    }
    $\bm{x}_{t+1} = {\rm SoftCorrection}(\bm{x}_{t+1}, \bm{x}_{t},\bm{x}_{t-1})$\\
}
\end{algorithm}

\section{Root Wrench Metric and Soft Correction}
\label{sec:soft}
While a passive ragdoll simulated in maximum coordinates with "$E_{mom} + \sum_i E_i$" adheres to physical validity, the introduction of $E_{kin}$ can introduce non-physical forces into the system. One non-physical behavior of our controlled character that cannot be regulated by $\{E_i\}$, is the fictitious "puppet-like" root force and torque (wrench) to unrealistically regain balance, ignoring the under-actuated nature of human movement.

Before we can attempt to minimize the use of such "magical forces", we need to develop a metric capable of quantifying them. To reduce unnatural balancing, this metric should ensure that the movement of the center of mass (CoM) is feasible via contact forces with the ground. Given that $E_{kin}$ can generate large non-physical forces and our contact model is very simplified, we cannot obtain meaningful contact forces directly from the simulation. Rather, we shall solve another optimization problem to find \textit{any} contact force that can explain, to the greatest extent possible, the current changes in CoM momentum. It is important to note that these contact forces are \textit{not} simulated within the system but are used solely as slack variables in this metric. We borrow such a metric from previous works \cite{Zheng2013, Zhang2021Manipnet}: 



\begin{equation}
\label{eq:manipNet-metric}
\min_{\bm{c}} \norm{\bm{F} \bm{c} - \bm{b}}^2, \textrm{~~~~subject~to~} \bm{c} \succeq \bm{0},
\end{equation}

Here, $\bm{F}$ is the contact force basis constructed from the Coulomb friction cone assumption, $\bm{c}$ are the unilateral contact force coefficients for each basis, and $\bm{b} = (\dot{\bm{P}}, \dot{\bm{L}})$ is the change in linear and angular momentum around CoM, estimated using finite-differencing and subtracting the effect from gravity and external environmental forces. The minimization is solved with non-negative least squares (NNLS). We direct readers to \cite{Zhang2021Manipnet} for technical details.

Once $\bm{c}$ is solved, we can interpret $\norm{\bm{b} - \bm{F} \bm{c}}^2$ as another quadratic energy term where $\bm{F} \bm{c}$ is constant ($\bm{F}$ can be assumed to only depend on $\bm{x}_t$ and not the $\bm{x}$ to be solved). Specifically, $\bm{b}$ is an affine function of $\bm{x}$, given that:

\begin{eqnarray}
   \dot{\bm{P}} &=& \sum_j m^{j} \bm{a}^{j} = \frac{1}{h}\sum_j m^{j} (\frac{\bm{x}^{j} - \bm{x}^{j}_{t}}{h} - \bm{v}^{j}_{t}), \label{eq:momentum} \\
    \dot{\bm{L}} &=& \sum_j m^{j} \bm{r}^{j} \times \bm{a}^{j} = \frac{1}{h} \sum_k m^{j} [\bm{r}^{j}]_{\times} (\frac{\bm{x}^{j} - \bm{x}^{j}_{t}}{h} - \bm{v}^{j}_{t}).
\end{eqnarray}


In these equations, $m^{j}$ represents the mass of the $j$-th joint particle, $\bm{a}^{j}$ denotes its acceleration, and $\bm{r}^{j}$ is its distance to the center of mass (CoM) location (moment arm), which remains constant if evaluated, similarly to $\bm{F}$, at the previous step $\bm{x}_t$. The skew-symmetric matrix is denoted by $[\bm{r}^{j}]_{\times}$. Here, central difference method is used for approximating $\bm{a}^{j}$. These equations enable us to write $\norm{\bm{b} - \bm{F} \bm{c}}^2$ succinctly as $\norm{\bm{B}\bm{x} - \bm{a} - \bm{F} \bm{c}}^2$, where $\bm{B} \in \mathbb{R}^{6 \times 3N}$ and $\bm{a} \in \mathbb{R}^{6}$ can be determined through manipulation of the equations above. 

One might be inclined to directly incorporate $\norm{\bm{B}\bm{x} - \bm{a} - \bm{F} \bm{c}}^2$ as an extra energy term in $E$ during projective dynamics iterations, where solving $\bm{c}$ using non-negative least squares (NNLS) would serve as the projection step. However, we found this approach to be unstable in practice. In early (e.g. first) projective dynamics iterations, the solved $\bm{c}$ can be significantly off when $\bm{x}$ is still far from the final solution. This could occasionally lead to an excessively large update in $\bm{x}$, given that there are infinitely many $\bm{x}$ that can minimize $\norm{\bm{B}\bm{x} - \bm{a} - \bm{F} \bm{c}}^2$. Instead, we choose to apply a post-processing correction, so we only evaluate $\bm{c}$ when $\bm{x}$ has stabilized to solution once all iterations are complete. Specifically, we solve for:


\begin{equation}
\label{eq:soft-correct}
    \bm{x}_{t+1}, \cdot = \argmin_{\bm{x}, \bm{\epsilon}} \norm{\bm{x} - \bm{x}_{pd}}^2 + w_p \norm{\bm{\epsilon}}^2, \textrm{~~~~subject~to~} \bm{B}\bm{x} - \bm{a} - \bm{F} \bm{c} = \bm{\epsilon},
\end{equation}
where $\bm{x}_{pd}$ represents the solution $\bm{x}$ obtained after all projective dynamics iterations, $w_p$ stands for the strength of the correction, and $\bm{\epsilon}$ is the slack variable for correction tolerance. $\bm{c}$ here is first solved from Eq. \ref{eq:manipNet-metric} by evaluating $\bm{b}$ on $\bm{x}_{pd}$ (i.e., $\bm{b} = \bm{B}\bm{x}_{pd} - \bm{a}$). Eq. \ref{eq:soft-correct} can be solved precisely by writing out the KKT matrix. We found that this approach is more stable than directly solving the unconstrained problem $\min_{\bm{x}} \norm{\bm{x} - \bm{x}_{pd}}^2 + w_p \norm{\bm{B}\bm{x} - \bm{a} - \bm{F}\bm{c}}^2$, which requires adding the low-rank $\bm{B}^T \bm{B}$ to the system matrix. 

It is crucial for such a correction to be gentle, since the root wrench metric $\norm{\bm{b} - \bm{F}\bm{c}}$ involves finite differencing for the second derivative and can be fairly imprecise, particularly for highly dynamic motions. We adjust $w_p$ to prevent an excessively large correction $\norm{\bm{x} - \bm{x}_{pd}}$. Conversely, to prevent minor corrections from accumulating over time and degrading the naturalness of the motion, we skip the post-correction when $\norm{\bm{b} - \bm{F}\bm{c}}$ is already small.

\section{Implementations}

\label{sec:impl}

In our algorithm, the kinematic energy term, $E_{kin}$, is based on a recent autoregressive generative model, HuMoR \cite{rempe2021humor}. HuMoR is trained on the large-scale AMASS dataset \cite{AMASS:ICCV:2019}, excluding motions involving terrains. When the next step joint positions $\bm{x}_{t+1}$ have been solved, the input to HuMoR is replaced with this solved position before querying HuMoR again for sampling the transition manifold at the next time step. HuMoR, along with many kinematic models, also considers joint rotations $\bm{q}$ as input, in addition to joint positions. Therefore, before querying HuMoR again, we need to determine joint rotations $\bm{q}_{t+1}$ from the solved joint positions $\bm{x}_{t+1}$. In this work, we adopt a swing-twist decomposition strategy \cite{li2021hybrik}, where the swing components of the joint rotations are completely determined by the bone vectors calculated from $\bm{x}$. As for the twist component, during the construction of the approximate manifold, the collection of samples $\mathcal{\hat{X}}_{t+1}$ are paired with corresponding $\mathcal{\hat{Q}}_{t+1}$ samples also predicted by the HuMoR model. We adopt the twist given by $\bm{q}_{kin}$, which is the corresponding $\bm{q}$ paired with the selected sample $\bm{x}_{kin}$ in the final projective dynamics iteration. For the leaf body nodes (head, hands, toes), whose rotations $\bm{x}$ does not inform, we directly prompt HuMoR to predict (so they might appear weird in out-of-distribution cases). We add a kinematic spring to damp leaf node rotations toward the identity (rest pose). Finally, although the HuMoR model also redundantly predicts joint Cartesian velocities $\bm{v}$ and autoregressively uses them, different from $\bm{x}$, we choose not to overwrite $\bm{v}$ using solved $\bm{x}$ with finite-differencing, as we found overwriting $\bm{v}$ can destabilize HuMoR's output.

For range-of-motion energy $E_3$, the VPoser model operates on $\bm{q}$ instead of $\bm{x}$, so we adopt the same swing-twist decomposition inverse kinematics. The VPoser model is a VAE with encoder represented as $\bm{z}= \bm{E}(\bm{q})$ and decoder $\bm{q} = \bm{D}(\bm{z})$. Following previous works \cite{rempe2021humor, weng2023diffusion}, $\norm{\bm{z}}$ can be considered as the validity of the pose, and we heuristically define $\norm{\bm{z}} < 6.0$ as the range of motion boundary. Thus, the projection $\bm{q}_r$ of a pose $\bm{q}$ back to the range of motion is as follows:
\begin{equation}
    \bm{q}_r =\begin{cases}
			\bm{D} ( ~\bm{E}(\bm{q}) / \norm{ \bm{E}(\bm{q})} \cdot 6.0 ~), & \text{if} \norm{ \bm{E}(\bm{q})} > 6.0, \\
            \bm{q}, & \text{otherwise},
		 \end{cases}
\end{equation}
after which we obtain $\bm{x}_r$ for Eq. \ref{eq:rom-energy} through forward kinematics.

Our current prototype does not utilize the detailed skinned geometry of SMPL for collision detection and contact resolution. Instead, we simply position manually-scaled spheres at the joints to approximate the collision shape. The weights for the energy terms do not have a significant impact on the overall quality of the motion. We set $w_1, w_2, w_3, w_{kin}$ to $1.0, 3.0, 3.0, 3.0$ respectively. We solve projective dynamics for three iterations at each simulation step. We simply set $\bm{M}$ to $\bm{I}$, assuming all joint mass particles have unit mass. While we can craft a more realistic $\bm{M}$, note that it will not affect the data-driven control term $E_{kin}$ since it operates on positions.

Our current unoptimized Python implementation of the method operates at 10 FPS, which is thrice slower than real-time, given that the system simulates with a time step of $1/30$ second. The performance bottleneck arises from the swing-twist IK (at each projective dynamics iteration for the RoM energy) and data transfer between CPU and GPU (for the neural-nets). Since our system matrix is small ($3N = 66$), optimized full-GPU code should most likely bring the performance beyond real-time. Notably, the sampling process for the kinematic energy can be easily batched on the GPU. Our code will be open-sourced upon publication.

\section{Evaluations}
Our experiments are designed to demonstrate three main objectives: (1) Our system's capability to generate a diverse array of responses to dynamic changes in a variety of motion categories; (2) The ease of motion editing and a variety of downstream applications enabled by our system; (3) The importance of the components proposed in our method.

All demonstrations in the supplementary video are generated by the existing HuMoR model combined with our human simulator. No additional motion control or planning are used, except for the modifications on energy terms and forces as detailed below. As such, the character’s responses and recovery motions to unexpected events are stochastically created due to the generative nature of the HuMoR model.

\subsection{Diverse Responses Across Various Motion Categories}

\begin{figure}[!h]
    \centering
    \includegraphics[width=\linewidth]{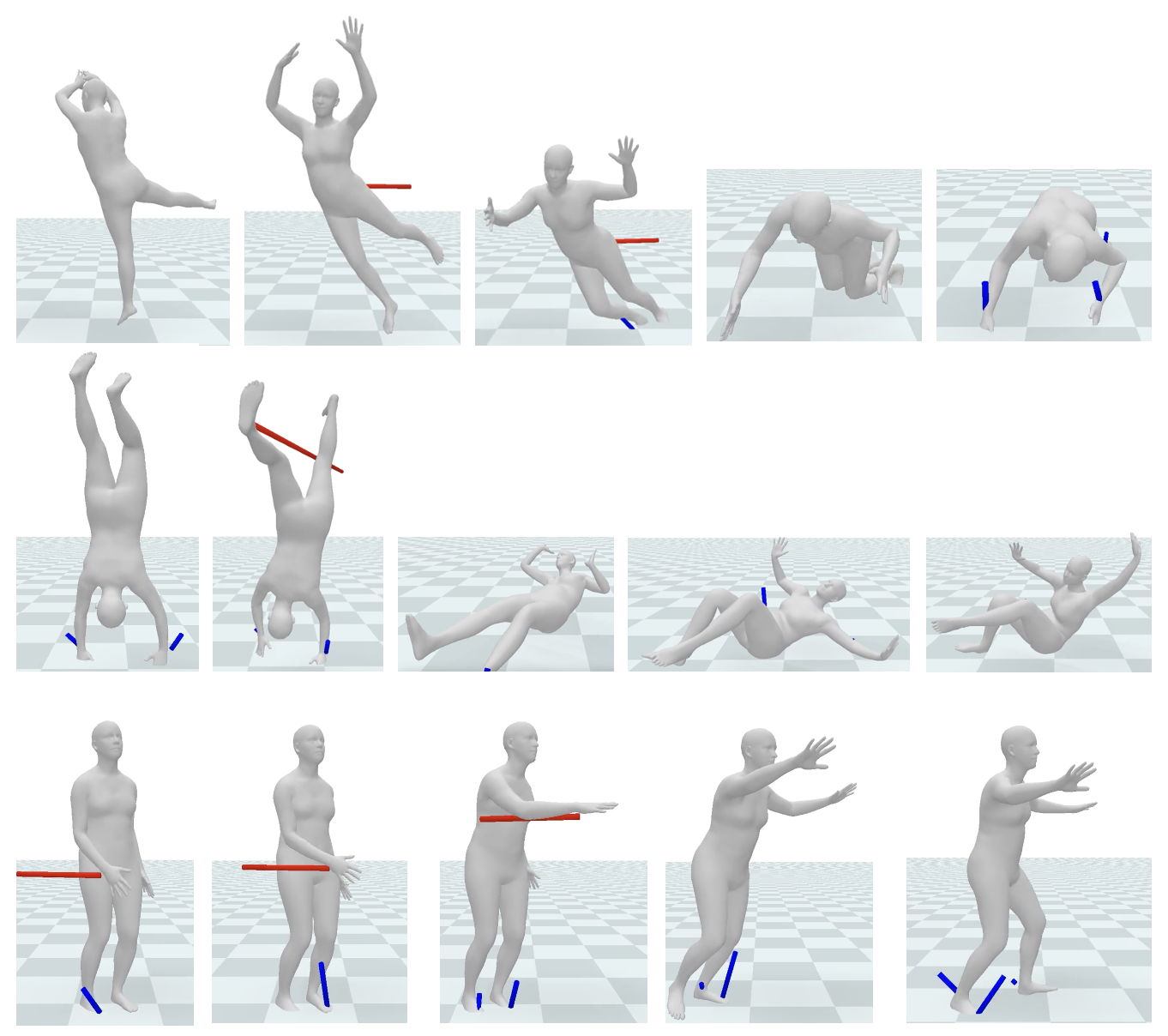}
    \caption{Our method can model responses to perturbations at different body parts. Red lines are pushing forces and blue lines represent contact forces calculated during root wrench correction.}
    \label{fig:forces}
\end{figure}

In the basic setting, the only environmental variation comes from an external perturbation force $\bm{f}$ (Eq. \ref{eq:dyn1}), which is applied either to the torso, foot, or the hand (Fig. \ref{fig:forces}). Our supplementary video illustrates the varied responses of our simulated human to both smaller (Video 1'01) and larger (Video 1'33) forces. We randomly initialize our system from poses within the CMU dataset \cite{cmuMocap} to encompass a broad range of motion categories. 

Our method can also model responses to perturbations when performing highly dynamic tasks, such as back flips (Video 5'15). In this specific demo, the system is asked to initially track a specific CMU motion until just before the perturbation occurs. In other words, initially, the system follows a mock motion prior that is deterministic and consistently transitions to the next frame in the MoCap database, only starting to track the actual motion prior (HuMoR) immediately after the onset of the external force. We employ this workaround since if we let HuMoR progresses randomly on its own without guidance, it tends to quickly converge to the motion categories that comprise the majority of its unstructured and unbalanced training data, i.e. locomotion.

\subsection{Downstream Motion Editing}
In our supplementary video, we showcase an array of downstream tasks that can be accomplished by our system in a straightforward manner.

\paragraph{Projectiles} We can enable our character to physically respond to projectile balls thrown at it by setting the external forces to be proportional to the depth of penetration between the projectiles and the collision spheres (Sec. \ref{sec:impl}) on the character body. This means that $\bm{f} = - \sum_j k \bm{d}_j$, where $\bm{d}_j$ is the penetration vector between the projectile sphere and the joint collision sphere, with this summed over all joints. The projectiles are then simulated according to Newton’s second law with $-\bm{f}$ (Video 0'29, 2'19).

\paragraph{Two-character collision} While it is possible to solve for two characters in one system, for ease of prototyping, we can also simulate each character independently, treating the other character’s body-collision spheres as if they were collidable objects in the environment. Since both characters are based off stochastic motion priors, a variety of interesting behaviors arise when two characters come closer to each other (Video 4'12). When approaching with a low speed, the characters try to dodge each other upon collision or attempt to taunt the other character out of way. When running fast into each other, we observe in one case where one character falls due to collision, but the other character regain balance and quickly runs away, leaving the first character struggling on the ground.

\paragraph{Direction following} Since the rollout of a motion prior will only generate random movements, previous generative kinematic models \cite{ling2020character} have shown that a high-level DRL policy can be trained to guide the kinematic generative model to follow a specific path or direction. With our framework, basic and greedy direction tracking can be achieved zero-shot by adding a center-of-mass-velocity energy, without the need for training an additional model. Specifically, we add the following energy to $E$:
\begin{equation}
    E_{com}(\bm{x}) = \norm{ \sum_j m^{j} \frac{\bm{x}^{j} - \bm{x}^{j}_t}{h}  - \bm{v}_{tar} \sum_j m^{j}}^2,
\end{equation}
where $\bm{v}_{tar}$ is the target velocity, $m^{j}$ and $\bm{x}^{j}$ are $j$-th joint's mass and position, same as in Eq. \ref{eq:momentum}. $E_{com}$ can also be considered as an additional soft physical constraint, that encourages $E_{kin}$ to project to poses in the transition manifold that align more closely with $\bm{v}_{tar}$. In the supplementary video (3'28), we make the character follow a moving red dot that changes the moving direction every 4 seconds. When the red dot moves at a slow speed ($\norm{\bm{v}_{tar}} = 1.2$ m/s), the character walks and turns accordingly. When we speed up the red dot ($\norm{\bm{v}_{tar}} = 2.5$ m/s), the character switches to running so it can catch up the target. Using the same technique, in the projectile demos (Video 0'29, 2'19), we also ask the character to attempt to dodge the projectile balls thrown at it by setting the direction of $\bm{v}_{tar}$ to be maximally distant to the directions of the balls.

\paragraph{Force-based constraints} One way to set positional constraints, such as a target in space for the right hand, is to add an external force $\bm{f}$ to the target joint (e.g., right hand), proportional to the constraint violation vector, pulling the target joint towards the position constraint. In the previous direction following demo (Video 3'28), when the character is asked to walk or run along a square trajectory (CoM tracking), we also add such a force-based constraint to its right hand, mimicking the character being led to walk or run. As shown in the video, this positional constraint is more closely satisfied at a lower center-of-mass speed (walking) and is violated more at a high speed (running). 


\paragraph{Energy-based constraints} We also experimented with setting positional constraints by adding a local energy term that penalizes the distance $E_{constr} = \norm{\bm{x}^{j} - \bm{x}^{j}_{tar}} ^ 2$ between the joint and its target position. In our video (1'58), we emulate a character getting tripped by immovable obstacles on the ground by fixing the heel joint for 1 second when it is about to touch an obstacle on the ground.

\paragraph{Tilting platform} Tilting the ground plane does not necessitate the addition of energy terms to the system. The character's changes in its behaviors are automatic due to unexpected collision with the ground. As the ground tilts, the character stops walking, crouches down, and sticks its arms out for balance (Video 2'41). This emergent behavior is entirely generated by the generative model, showcasing HuMoR's inherent generalization ability as its training data only contains motions on a flat ground.


\paragraph{Stiff knee} Our method is also capable of modeling dynamics variations intrinsic to the character's skeleton model. By adding a strong spring between the left hip joint and the left heel joint, similar to the ones we used to model bones, we can simulate a walking motion with a stiff knee (Video 4'03).

\subsection{Ablation Study}

To demonstrate the significance of the components proposed in our method for generating realistic response motions, we performed the following ablation studies (Video 4'41). Without HuMoR, the character would behave like a passive rag doll without control. Without the range of motion energy $E_3$, the system is prone to generating unnatural poses that fall outside of the typical human range of motion. Without the collision energy term $E_2$, the character might interpenetrate with the ground during perturbations and recoveries. Without the soft correction for root wrench, the character uses puppet-like ghost force to recover from large perturbations without utilizing contact.

\subsection{Analysis on the Energy Trajectory}
\label{sec:analysis-root}
\begin{figure*}
    \centering
    \includegraphics[width=0.9\linewidth]{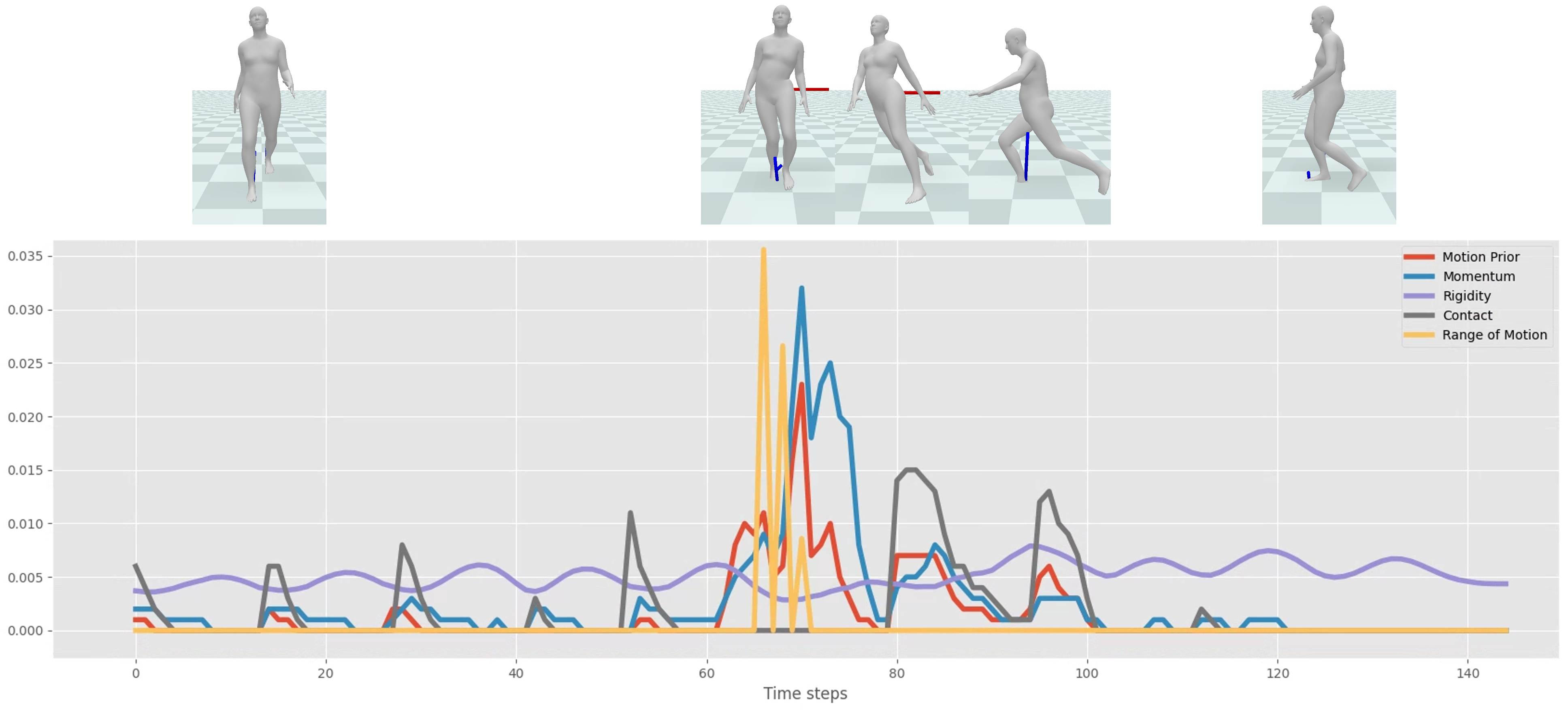}
    \caption{Contributions of each energy term during synthesizing a motion of perturbed walking. External force happens at frame 63. At each step, the solve state compromises between motion prior and other energies, regulating the recovery of motion prior back to training data distribution to be physically valid.}
    \label{fig:energies-one-motion}
\end{figure*}

Fig. \ref{fig:energies-one-motion} displays a synthesized motion wherein during normal walking, the character is perturbed by an external force at the pelvis (represented by the red line in the rendering), takes a few recovery steps, and returns to normal walking. During this trajectory, we note that at the onset of perturbation, the RoM and Momentum energies experience increased violation, making the solved motion to deviate from the distribution of motion prior and driving $E_{kin}$ to increase as well. Subsequently, during the recovery steps, $E_{kin}$ remains relatively high, suggesting that the motion prior has not yet returned to the AMASS training distribution. The Contact energy is high here to prevent the recovery steps from appearing implausible (i.e., penetrating the floor too much). 

\begin{figure}
    \centering
    \includegraphics[width=\linewidth]{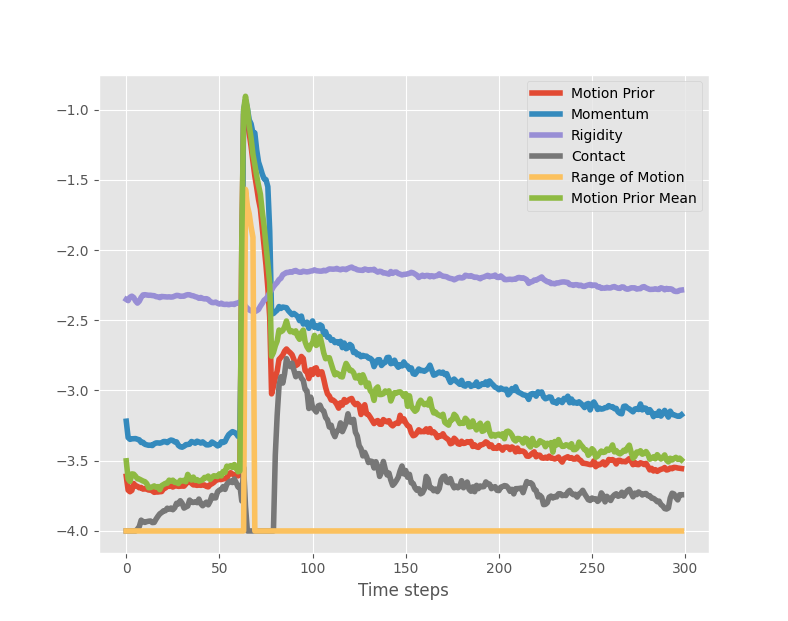}
    \caption{Median trajectory of different energy terms in our method, across 250 random motions all being perturbed from steps (frames) 63 to 78. Our method guides the motions to return to training data distribution over time, as evidenced by decreasing motion prior energy.}
    \label{fig:energies-250-motions-ours}
\end{figure}

To illustrate that our system can guide the motion prior's recovery back towards the original training data distribution, we extend Fig. \ref{fig:energies-one-motion} to analyzing 250 random motions, all perturbed at the same point of time in the trajectory (Fig. \ref{fig:energies-250-motions-ours}). A logarithmic y-scale is used for visual clarity, and the medians per frame across the 250 motions are plotted. Similar to Fig. \ref{fig:energies-one-motion}, Fig. \ref{fig:energies-250-motions-ours} shows that during the unperturbed stage, the kinematic energy largely aligns with physics-based energy terms, and hence all energy levels are low. Both kinematic and other energy terms increase at the point of perturbation, causing the motion to deviate from the motion prior as a compromise between the energies. The median energies gradually decrease, reflecting that it takes time for motions to recover to training distribution (e.g. get up and resume normal walking). We also plot a hypothetical energy term, "motion prior mean" (green line in Fig. \ref{fig:energies-250-motions-ours}), which represents the kinematic energy level if $E_{kin}$ was calculated from a single-point manifold, as given by the mean/mode of the HuMoR's decoder, instead of having access to a full manifold. The results demonstrate that, especially during recovery (around frames 100 to 150), using the full manifold reduces $E_{kin}$ by at least 30\% (note the logarithmic scale), indicating that the system prefers having the flexibility of choosing from a manifold over a single point.

\begin{figure}
    \centering
    \includegraphics[width=\linewidth]{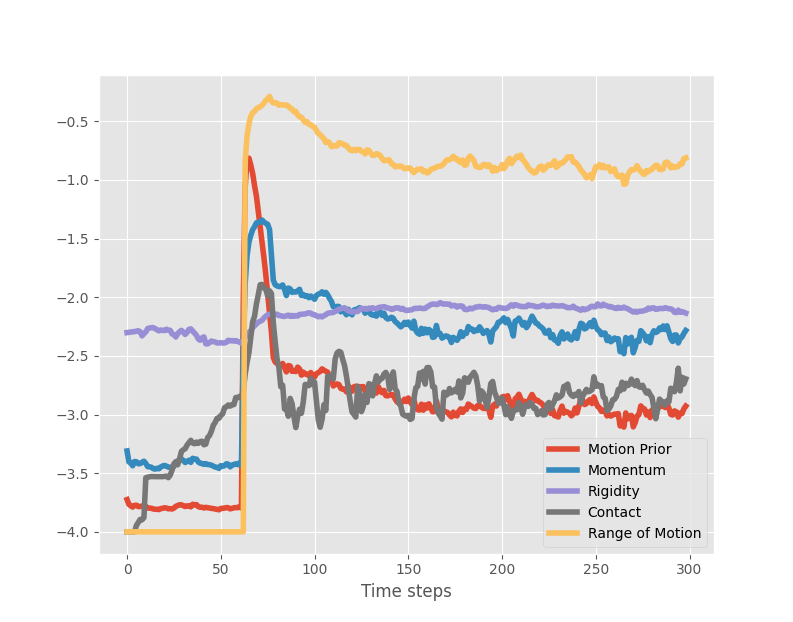}
    \caption{Median trajectory of different energy terms in an ablation method without RoM and contact energies being used in simulation, across 250 random motions all being perturbed from steps (frames) 63 to 78. Final energies of both motion prior term and the physics terms are at least 5x higher than our full method, and show no signs of further decrease.}
    \label{fig:energies-250-motions-ablation}
\end{figure}

Finally, for an ablation system without the Contact and RoM energy terms (we retain the Rigidity energy, effectively allowing the generalized coordinate representation to always hold), Fig. \ref{fig:energies-250-motions-ablation} shows that both kinematic and physics-based energies are at least 5 times higher at the end of the trajectory and do not demonstrate a trend towards final recovery.

\subsection{Analysis on Root Wrench Metric and Correction}

\begin{figure*}
    \centering
    \includegraphics[width=0.9\linewidth]{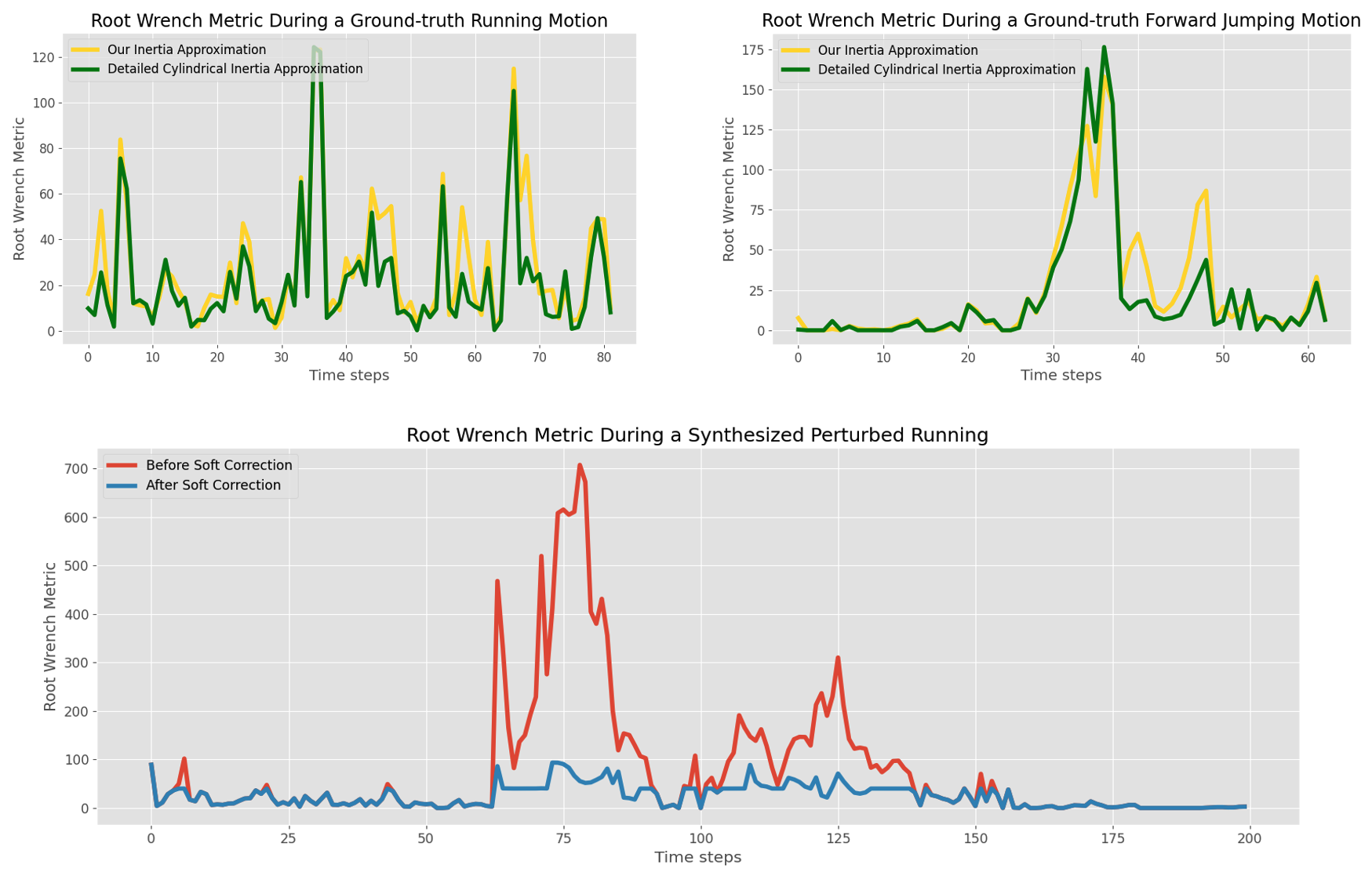}
    \caption{Up: Root wrench usage metric is non-negligible even during ground-truth MoCap trajectories, shown running (left) and jumping (right) motions. Bottom: Our soft correction technique aims to bring the metric down to a similar range as ground-truth motions.}
    \label{fig:root-metric}
\end{figure*}

In Sec. \ref{sec:soft}, we introduced a \textit{soft} correction mechanism to regulate the usage of the "root magic force". Given that the measurement of this metric involves twice finite differencing, it can be numerically sensitive. Fig. \ref{fig:root-metric} Top provides evidence that even for ground-truth mocap motions, such as jumping and running, the root wrench metric can have a non-negligible value. Intuitively, even a small inaccuracy in MoCap data can have a substantial impact on the metric, because finite differencing involves multiplying by $1/h^2$.


We also demonstrate that our simple inertia approximation using mass points at joints does not significantly affect the estimation of the root wrench metric. This can be observed by comparing it with a more detailed inertia approximation that treats all body parts as manually-scaled cylinders (scaled to roughly match the skinned SMPL body shape \cite{SMPL:2015}). As can be seen in Fig. \ref{fig:root-metric} Top-left and Top-right, the cylindrical approximation yields similar root wrench metric for ground-truth motions. 

In Fig. \ref{fig:root-metric} Bottom, we present an illustrative example of how our soft correction method reduces the root wrench metric for a synthesized motion of perturbed running. During this motion, the red line represents what the root metric would have been without our correction mechanism, while the blue line shows the actual value for this motion, post-correction. Our method only brings down the metric to a similar range as seen in ground-truth motions, instead of completely eliminating it, as the metric is not exact. As such, some motions that visually use non-physical root forces, but has the metric in range, will not be corrected (Sec. \ref{sec:limitations}).



\subsection{Comparisons with Reinforcement-Learning-Based Methods}
We explore an alternative framework to generate physics-based human motion without using deep reinforcement learning (DRL). By decoupling the training to be only kinematics, our method scales to large datasets much more easily and with less compute. In the accompanying video (0'48, 1'20), we present a few non-rigorous qualitative comparisons to the state-of-the-art DRL method ASE \cite{Peng2022ASE}. We focus on the responses of the two methods to small and large external forces, since when without perturbation our method generates motions with quality on-par to HuMoR. Though ASE is trained on a much smaller dataset which specifically contains perturb and get-up motions, our method can generate more compliant and realistic (in terms of physical capabilities of real humans) responses. Our method also generates more diverse responses thanks to being trained on the large-scale AMASS data. On the other hand, motions from ASE can sometimes be more physical than ours since it is trained in a full-blown physics engine.

\subsection{Failure Cases}

It is challenging to define failure quantitatively for this work, as the kinematic model will continue to drive the character towards random poses even if it enters a unstable state. Also, with our physics-based energies in similar forms with common physicality metrics, quantitative physical metrics could appear valid even in obvious failure cases from visual inspection. Our video (5'43) showcases several such qualitative failure cases, where the most common issue can be characterized as "super-human behavior". For example, the character could quickly spin near the floor while lying. In such cases, the motion prior clearly enters a disastrous state from which it cannot recover even with guidance of the system. However, since the character still makes intermittent contact with the floor, the motion would be deemed valid by the soft correction routine. This behavior suggests that additional physical constraints, such as torque limits, may need to be added to the system, since such fast spinning indicates an effectively huge amount of control torque being added to the system. We provide more discussion on current limitations in Sec. \ref{sec:limitations}.

\section{Discussion}

This paper introduces a minimal human physics simulator based on projective dynamics, that can be plugged into a learned generative kinematics motion prior to augment it with dynamic capabilities, without any further training beyond the pretrained motion prior. Thanks to the motion diversity inherited from the motion prior, simulating various dynamics responses to physical and environmental changes is surprisingly easy.

\subsection{Alternative Formulations} \label{sec:alternative} Our current framework poses no constraint on the type of generative motion models, as long as they can be sampled to generate a set of next states for manifold approximation. We note that Eq. (10) cannot be used in vanilla optimization-based simulators as an energy term, due to construction of $\mathcal{X}_{t+1}$ involving non-differentiable sampling. On the other hand, projective dynamics does not require an analytical or differentiable expression of Eq. (10). 

It’s technically also feasible to train an Energy-Based Model \cite{lecun2006tutorial}, explicitly predicting the scalar distance (score) to training data distribution, given any motion transition as input. An EBM, being a closed-form differentiable neural network, can theoretically work with any energy-based simulation framework. In fact, that is precisely what we experimented with early on in this work before deciding to switch to projective dynamics - despite recent advances in training EBMs, we found they can still be unstable to train and hard to scale to a large dataset. In contrast, projective dynamics frees us to use any motion prior that scales well, such as VAEs \cite{rempe2021humor}, discretized token-based Transformers \cite{li2020learning}, or more recent Diffusion Models \cite{tevet2023human}.

\subsection{Limitations} \label{sec:limitations} Due to the sensitivity of the root wrench metric (Sec. \ref{sec:analysis-root}), we choose to allow a small amount of root wrench, which may occasionally lead to puppet-like motions. Further research is required to base this metric on a time window rather than a single step, which may potentially mitigate its sensitivity resulted from finite differencing. The root wrench correction should also ideally be incorporated into the projective dynamics solver rather than as post-processing. Particularly at states where data coverage is poor, the motion prior could mandate a very unrealistic transition causing the root correction to solve for a very large correction step in response, which might violate previously solved energy terms causing artifacts such as foot sliding. 

We note that our foot sliding or penetration metric (equivalently to the level of contact energy, Fig. \ref{fig:energies-250-motions-ours} and \ref{fig:energies-one-motion}) and physicality metric (equivalent to root wrench, Fig. \ref{fig:root-metric}) should be at least as good as HuMoR on unperturbed AMASS training motions, in which case our method degenerates to physics-based post-processing on top of kinematic models. During perturbation recoveries, our physics-based energy values become higher, which equivalently indicates quantitative physics-based metrics becoming worse.

As our method involves no dependency on existing physics engines or libraries, our current implementation from scratch involves several simplifications. Our method does not simulate toes, hands, or head, and only uses spheres at each joint as collision geometries. As result, our visualization (using fully skinned SMPL model) inevitably sees occasional foot and hand penetration, as well as self-penetration between body parts, even though our RoM energy provides some implicit mitigation to self-penetration. We also use a maximal-coordinate stick figure to ease the incorporation of projective dynamics, with the twist motion excluded in the projective dynamics solver and fully predicted kinematically from HuMoR. Extending the framework to articulated rigid bodies would provide a better representation.

\begin{acks}

This work was supported by the Wu Tsai Human Performance Alliance at Stanford University, the Stanford Institute for Human-Centered AI (HAI), Meta Reality Labs, and NSF:CCRI 2120095. Jungdam Won was partially supported by the New Faculty Startup Fund from Seoul National University and the ICT (Institute of Computer Technology) at Seoul National University. We would like to thank Seunghwan Lee, Andrew Spielburg, Tom Van Wouwe, Jackson Kuan-Chieh Wang, Albert Wu, Eris Jiayi Zhang, Alexander Winkler, Deepak Gopinath, and anonymous reviewers for helpful discussions.

\end{acks}

\bibliographystyle{ACM-Reference-Format}
\bibliography{main}

\end{document}